\documentclass[twocolumn,amsmath,amssymb]{revtex4}

\def\gsim{ \lower .75ex \hbox{$\sim$} \llap{\raise .27ex \hbox{$>$}} }

\def\lsim{ \lower .75ex \hbox{$\sim$} \llap{\raise .27ex \hbox{$<$}} }

\def\IZ{\relax\ifmmode\mathchoice
{\hbox{\cmss Z\kern-.4em Z}}{\hbox{\cmss Z\kern-.4em Z}}
{\lower.9pt\hbox{\cmsss Z\kern-.4em Z}} {\lower1.2pt\hbox{\cmsss
Z\kern-.4em Z}}\else{\cmss Z\kern-.4em Z}\fi}
\def\IR{\relax{\rm I\kern-.18em R}}

\def\one{{\hbox{ 1\kern-.8mm l}}}

\def\bm{\bibitem}

\newlength{\bredde}
\def\slash#1{\settowidth{\bredde}{$#1$}\ifmmode\,\raisebox{.15ex}{/}
\hspace*{-\bredde} #1\else$\,\raisebox{.15ex}{/}\hspace*{-\bredde}
#1$\fi}
\newcommand{\ft}[2]{{\textstyle\frac{#1}{#2}}}

\newsavebox{\zzzbar}
\sbox{\zzzbar}
  {\setlength{\unitlength}{0.9em}
  \begin{picture}(0.6,0.7)
  \thinlines
  \put(0,0){\line(1,0){0.6}}
  \put(0,0.75){\line(1,0){0.575}}
  \multiput(0,0)(0.0125,0.025){30}{\rule{0.3pt}{0.3pt}}
  \multiput(0.2,0)(0.0125,0.025){30}{\rule{0.3pt}{0.3pt}}
  \put(0,0.75){\line(0,-1){0.15}}
  \put(0.015,0.75){\line(0,-1){0.1}}
  \put(0.03,0.75){\line(0,-1){0.075}}
  \put(0.045,0.75){\line(0,-1){0.05}}
  \put(0.05,0.75){\line(0,-1){0.025}}
  \put(0.6,0){\line(0,1){0.15}}
  \put(0.585,0){\line(0,1){0.1}}
  \put(0.57,0){\line(0,1){0.075}}
  \put(0.555,0){\line(0,1){0.05}}
  \put(0.55,0){\line(0,1){0.025}}
  \end{picture}}

\newcommand{\ena}{\end{eqnarray}}
\newcommand{\beqa}{\begin{eqnarray}}
\newcommand{\eeqa}{\end{eqnarray}}
\newcommand{\bea}{\begin{eqnarray}}
\newcommand{\eea}{\end{eqnarray}}

\newcommand{\be}{\begin{equation}}
\newcommand{\ee}{\end{equation}}
\def\ft#1#2{{\textstyle{\frac{\scriptstyle #1}{\scriptstyle #2} } }}
\def\fft#1#2{{\frac{#1}{#2}}}

\usepackage{graphicx}

\def\im{{{\rm i}}}
\def\nn{\nonumber}
\def\ben{\begin{equation}}
\def\een{\end{equation}}

\def\bea{\begin{eqnarray}}
\def\eea{\end{eqnarray}}
\def\be{\begin{equation}}
\def\ee{\end{equation}}
\def\beq{\begin{eqnarray}}
\def\eeq{\end{eqnarray}}

\def\({\left (}
\def\){\right )}
\def\[{\left [}
\def\[{\right ]}

\def\ba{\begin{eqnarray}}
\def\ea{\end{eqnarray}}

\def\del{{\partial}}

\input amssym.def
\input amssym.tex

\usepackage{dcolumn}

\begin{document}

\rightline
\hfill {UPR-1220-T\ \ \ 
DAMTP-2010-87\ \ \
MIFPA-10-46\ \ \
}

\title{Universal Area Product Formulae for Rotating and Charged Black Holes
in Four and Higher Dimensions}

\author{\bf M.Cveti\v c,$^{1,4}$ G.W. Gibbons,$^2$ C.N. Pope$^{2,3}$}

\affiliation{
\vskip .5mm
{\it$^1$  Department of Physics and Astronomy,
University of Pennsylvania, Philadelphia, PA 19104, USA}  \vskip .5mm              
{\it $^2$ DAMTP, Centre for Mathematical Sciences, Cambridge University, Wilberforce Road, Cambridge, CB3 0WA, UK }
\vskip .5mm
{\it $^3$ George P. \& Cynthia W. Mitchell Institute for Fundamental Physics and Astronomy, Texas A\&M University, College Station, TX 77843-4242, USA}
\vskip .5mm
{\it$^4$ Center for Applied Mathematics and Theoretical Physics,
University of Maribor, Maribor, Slovenia}\vskip 0.5mm}

\begin{abstract} 
 We present explicit results for the product of  all horizon areas for  
general rotating multi-charge black holes, both in asymptotically flat 
and asymptotically anti-de Sitter spacetimes in four and higher dimensions.  
The expressions  are universal, and  depend only on the 
quantized charges, quantized angular momenta and  the cosmological constant. 
If  the latter is also quantized these  universal results may provide a 
``looking glass''   for  probing the  microscopics of general black holes.
 
\end{abstract}


\maketitle


Explaining the origin of the black hole entropy
$S=\ft14 A$
at the microscopic level, where $A$ is the area of the outer event horizon,
is an outstanding problem for quantum theories of gravity. Significant 
insights have been achieved for supersymmetric, 
asymptotically flat, multi-charged black holes  in four and five dimensions 
\cite{StromingerVafa},  where the  microscopic degrees of freedom 
can be explained in  terms of  a two-dimensional conformal field theory.  
More recent work has focused on  the microscopic  
entropy of extreme rotating solutions \cite{Stromingeretal}.  By contrast, 
the detailed microscopic origin of the entropy  of {\it non-extremal} rotating 
charged black holes  remains an open problem, although recently there has 
been some promising progress \cite{CastroMaloneyStrominger}.

    Greybody factors (i.e. absorption coefficients) 
and radiation spectra  provide another approach to 
probing  the black hole structure. An intriguing property of  multi-charged 
rotating  black holes (in maximally supersymmetric supergravity theories) 
is  that their wave equations are separable.  The radial equation 
has poles at the locations of the horizons, where the 
radial component of the metric degenerates, with residues proportional to 
the inverse squares of the surface gravities, and so the  Green functions 
are  sensitive to the geometry near {\it all} the black hole horizons, and not 
just the outermost one. 
The thermodynamic properties, including the
 surface gravity and area  at each horizon, can therefore 
be expected to  
play a role in understanding the entropy at the  microscopic level.  

   Some of these  ideas have been explored  for asymptotically flat, 
rotating,  multi-charged black holes in
four and five spacetime dimensions. (Explicit solutions  were given in 
\cite{CYI,CYII}, as generating solutions of maximally supersymmetric 
${\cal N}=4$ (or ${\cal N}=8$) supergravities, obtained as  
toroidal compactifications of the heterotic string 
(or of Type IIA string or M-theory).)  In addition to their mass $M$,  
in four dimensions these solutions  are specified by  four charges  
$Q_i$ ($i=1,\cdots ,4$) and one angular momentum $J$,  and in five dimensions 
by three charges $Q_i$ ($i=1,2,3$) and two angular momenta $J_{1,2}$. 
These black holes have just two horizons, and 
the area of the 
outer horizon has the tantalizing form \cite{CYI}
\ben
S_+= 2\pi(\sqrt{N_L} + \sqrt{ N_R})  \,.\label{Splus}
\een
where the integers $N_L$ and $N_R$ may be viewed as the excitation numbers 
of the 
left and right moving modes of a weakly-coupled two-dimensional conformal
field theory.  $N_L$ and $N_R$ depend explicitly on all 
the black hole parameters. 
It was  pointed out, first  in  the static case 
\cite{Larsen} and later for the general rotating black holes 
\cite{CLI,CLII}, that the entropy of the inner horizon, $S_-=\ft14A_-$,
is
\ben
S_-= 2\pi(\sqrt{N_L}- \sqrt{N_R})   \,. 
\een
 From this and (\ref{Splus}), it follows that the product of the 
inner and outer horizon 
entropies satisfies $S_+ S_-= 4\pi^2(N_L-N_R$),
which in terms of the underlying conformal field theory would be 
interpreted in terms of a level-matching condition. $S_+ S_-$ 
should therefore also
be an integer \cite{Larsen, CLI,CLII}. (This point was recently 
re-emphasized  in 
\cite{CLIII}.) It was found that $S_+ S_-$ is indeed 
quantized, and intriguingly, it is expressed solely in terms of the 
quantized charges and 
quantized angular momenta. In particular, it is modulus-independent,
taking the forms
\bea
&& \!\!\!\!\!\!\!\!\!\!\!S_+S_- = 4\pi^2 (\prod_{i=1}^4 Q_i+ J^2)\label{4res}\\
&& \!\!\!\!\!\!\!\!\!\!\! S_+S_- \!\! =\! 
   4\pi^2(\prod_{i=1}^3 Q_i + J_R^2-J_L^2)
\!=\! 4\pi^2(\prod_{i=1}^3 Q_i + J_a J_b )\label{5res}
\eea
in four and five dimensions respectively. 
(These results were implicit in 
\cite{CLI,CLII}, though not explicitly  evaluated.)  The solutions considered
here can be viewed as ``seed solutions'' from which the complete
families can be generated.  The expressions for $S_+ S_-$ would be expressed
in terms of S-, T- and U-duality invariants built from the charges in the
general case. 

   In a parallel development, Ansorg and collaborators
 \cite{Ansorg1,Ansorg2,Ansorg3,Ansorg4,Ansorg5,Ansorg6,Ansorg7}
studied general axisymmetric stationary solutions of 
Einstein-Maxwell theory in four 
dimensions, with sources external to the horizons.  They obtained  
striking ``universal''  formulae
expressing
the areas  $A_{\pm}$ of the outer  and inner  Killing horizons  
in terms of  
the total angular momentum $J$ and total charge $Q$. In particular,
for  Kerr-Newman  black holes, they found (in the normalisation 
conventions we use in the remainder of this paper)
\ben
A_+^2 \le A_+A_- = ( 8\pi J)^2 + (4 \pi Q^2)^2  \,,
\label{extreme} 
\een 
in agreement (after conversion to  our conventions)
with the result given above in the special case that the
four charges are set equal.
Note the inequality (\ref{extreme}) may be interpreted 
as  a general criterion
for extremality, and has been used to prove a No-Go theorem for the possibility
of force balance between two rotating black holes \cite{Hennig}.
 
   It is natural to enquire whether analogous properties hold for
more general classes of black holes; and especially, for those where
the radial metric function has more than two zeroes. Examples
include charged or rotating black holes in four or five dimensional gauged 
supergravity, and in more than five dimensions with or without gauging.
The wave equations in 
these backgrounds 
will have dominant contributions associated with poles at each of these 
zeroes. One can therefore again expect that the thermodynamics associated 
with {\it each} pole will play a role in governing the properties of the
black hole at the microscopic level.  
At event horizons or Cauchy horizons, the metric at
fixed radius has signature $(0,+,+,\cdots,+)$; that is, it describes a null
hypersurface.  However, it may happen that the induced metric has 
signature $(0,-,+,+,\cdots, +)$; in other words that the hypersurface is
time-like, and the area of this ``pseudo-horizon'' \cite{time} 
is pure imaginary. The metric radial function may also have zeroes for
complex values of the radial variable; these occur in conjugate pairs.
In what follows, we shall
just refer to zeroes of the radial function as horizons, 
regardless of whether the areas are real, imaginary or complex. 

   If it is indeed the case that geometries near all the horizons 
are involved in governing the microscopic behaviour of the black hole,
one might expect that the formulae (\ref{4res}) and (\ref{5res}) should
generalise, for the more general black hole examples, to expressions involving 
the products of {\it all} the horizon entropies or areas.  This would
suggest the possibility of an 
explanation for the microscopic behaviour of such black holes in terms
of a field theory in more than two dimensions.  

   We shall present results for the products of horizon areas in examples
that include
certain rotating black hole solutions in gauged supergravities in dimensions
4, 5, 6 and 7, and also Kerr-anti de Sitter rotating black holes
in arbitrary spacetime dimensions.  For the sake of brevity, we shall not
present the details of our calculations in all cases, and 
instead, we have selected one example, namely 
the rotating black hole in five-dimensional minimal gauged
supergravity, for which we present the calculation of the area-product formula
in more detail.

   The formulae that we obtain for the area products are  universal; they
  depend only on quantized charges, quantized angular momenta and  
the cosmological, or gauge-coupling, constant. In the case that the latter 
is also quantized  (such as arises in compactifications of string theory,
as discussed, for example, in \cite{Bousso}), these results are indeed 
suggestive of some underlying microscopics. 
For example, one may 
speculate that asymptotically 
anti-de Sitter black holes in four and five dimensions, for which there
are three horizons, may have a 
microscopic origin in three-dimensional Chern-Simons theory.

We shall use normalisation conventions where the Lagrangian density for
gravity and Maxwell field(s) is of the form 
\be
{\cal L}= \fft{1}{16\pi G}\, \Big( R - \sum_i \Phi_i(\phi) F^i_{\mu\nu} 
 F^{i\, \mu\nu}+ (D-1)(D-2) g^2\Big)\,,
\ee
where the functions of scalar fields (if present) are such that
$\Phi^i(\phi)$ tends to unity at infinity for the black-hole solutions.
We define charge(s) and angular momenta by
\be
Q_i= \fft1{4\pi}\int \Phi^i(\phi) {*F^i}\,,\qquad
J_i=\fft1{16\pi} \int {*dK^i}\,,
\ee
where $K^i=K_\mu^i dx^\mu$ and $K^{i\, \mu}\del_\mu=\del/\del\psi_i$,
where $\psi^i$ is the azimuthal coordinate, with period $2\pi$, in the 
2-plane associated with the angular momentum $J_i$.

  Our results for the products of the horizon areas for rotating
black holes in gauged supergravities in dimensions 4, 5, 6, and 7 are
as follows:
\medskip

\noindent
{\bf $D=4$ ungauged 4-charge \cite{CYI}:} 
\smallskip

$A_+ A_- = (8\pi J_a) (8\pi J_b) + 
                     256\pi^2 \prod_{i=1}^4 Q_i$,

\bigskip
\noindent
{\bf $D=4$ gauged pairwise equal charges \cite{d4gaugebh}:}

\smallskip
$\prod_{\alpha=1}^4 A_\alpha = (4\pi)^2\, g^{-4}\, (8\pi J)^2 +
   4 g^{-4}\, (4\pi Q_1)^2\, (4\pi Q_2)^2$.

\bigskip
\noindent
{\bf $D=5$ ungauged 3-charge \cite{CYII}:}
\smallskip

$A_+ A_- = (8\pi J_a) (8\pi J_b) + 256\pi \prod_{i=1}^3 Q_i$,

\bigskip
\noindent
{\bf $D=5$ minimal gauged \cite{d5gaugemin}:}
\smallskip

$\prod_{\alpha=0}^2 A_\alpha = -2 \im\, \pi^2\, g^{-3}\, (8\pi J_a)(8\pi J_b)
    -\im\, g^{-3}\, \Big(\fft{8\pi Q}{\sqrt3}\Big)^3$, \label{d5gauge}

\bigskip
\noindent
{\bf $D=5$ gauged $Q_1=Q_2\ne Q_3$ \cite{d5gaugemei}:}
\smallskip

$\prod_{\alpha=0}^3 A_\alpha =
-\ft{2\im\, \pi^2}{g^3}\, (8\pi J_a)(8\pi J_b) 
 - \ft{\im}{g^3}\, (8\pi Q_1)^2\, (8\pi Q_3)$.

\bigskip
\noindent
{\bf $D=6$ gauged \cite{chow6}:}
\smallskip

$\prod_{\alpha=1}^6 A_\alpha  = g^{-8}\, \Big(\fft{8\pi^2}{3}\Big)^2\, 
(8\pi J_a)^2\, (8\pi J_b)^2
   + g^{-6}\, \Big(\fft{8\pi Q}{3}\Big)^6$.

\bigskip
\noindent
{\bf $D=7$ gauged \cite{chow7}:}
\smallskip

$\prod_{\alpha=1}^4 A_\alpha = \pi^3\, g^{-5}\, \prod_{i=1}^3 
(8\pi J_i) - g^{-4}\, (2\pi Q)^4$.

\bigskip

   Note that we have included the cases of the 4-charge $D=4$, and the
3-charge $D=5$, solutions in ungauged supergravities, which were already 
presented as entropy-product formulae in the introduction.  This is done
for the sake of uniformity, using the normalisation conventions that we 
follow in the rest of the body of the paper. The citation in each heading
above refers to the paper where the black hole solution was constructed.

  To illustrate how these calculations may be performed, we
shall present the example of the rotating black hole in five-dimensionsal
minimal gauged supergravity. The horizons are located 
at the roots of the radial function 
\be
\Delta(r)=(1+g^2 r^2)(r^2+a^2)(r^2+b^2) +q^2+2 a b q -2m r^2
\label{5h}
\ee
that appears in the metric found in \cite{d5gaugemin}.  This is a 
cubic polynomial in $r^2$, and so there are six roots in total, occurring in 
pairs for which $r^2$ takes the same value.  We may view $x=r^2$ as the
radial variable, and thus just consider 3 roots.  We may write $\Delta$ as
\be
\Delta(r)=g^2\prod_{\alpha=0}^2 (r^2-r_\alpha^2)\,.\label{3roots}
\ee
The horizon areas are
\be
A_\alpha= \fft{2\pi^2[(r_\alpha^2+a^2)(r_\alpha^2+b^2) +abq]}{
               \Xi_a\Xi_b\, r_\alpha}\,.
\ee
Using (\ref{5h}) and $\Delta(r_\alpha)=0$, we can write this as
\be
A_\alpha= -\fft{2\pi^2\, (2m+abq g^2)}{\Xi_a\Xi_b\, (1+g^2 r_\alpha^2)
r_\alpha}\, \Big[\fft{q(q+ab)}{2m+abq g^2} -r_\alpha^2\Big]\,.\label{As}
\ee
Noting from (\ref{5h}) and (\ref{3roots}) that we may write $\prod_\alpha
(c^2-r_\alpha^2)$ as $g^{-2}\, \Delta(c)$, for any $c$, 
it is then straightforward to 
evaluate the product of the $A_\alpha$.  With the angular momenta and
the charge given in terms of the rotation parameters $a$ and $b$, the
mass parameter $m$, and the charge parameter $q$ by \cite{d5gaugemin} 
\bea
J_a &=& \fft{\pi\,[2am + q b\,(1+g^2 a^2)]}{4\Xi_a^2\, \Xi_b}\,,\\
J_b &=& \fft{\pi\,[2bm + q a\,(1+g^2 b^2)]}{4\Xi_b^2\, \Xi_a}\,,\quad
Q = \fft{\sqrt3\, \pi\, q}{4\Xi_a\Xi_b}\,,\nn
\eea
where $\Xi_a=1-a^2 g^2$ and $\Xi_b=1-b^2 g^2$,
a straightforward calculation then gives the result we listed above.  The
calculations for the other examples can be performed in a similar manner.

   For the Kerr-AdS metrics in arbitrary dimensions \cite{gilupapo1,gilupapo2},
it is necessary to separate the cases of even dimensions, $D=2N+2$, and odd
dimensions, $D=2N+1$.  In each case there are $2N+2$ horizons and 
$N$ angular momenta $J_i$.  When $D=2N+1$, the radial metric function is
a function of $r^2$, and the product over all horizons is equivalently 
expressible as the square of the product over just $N+1$ horizons corresponding
to a single choice of square root for each $r_\alpha^2$.  Our results
for the horizon area products in $D$-dimensional Kerr-AdS are 
\bea
D=2N+2:&& \prod_{\alpha=1}^{2N+2}A_\alpha = g^{-4N}\, ({\cal A}_{D-2})^2\,
   \prod_{i=1}^N (8\pi J_i)^2\,,\nn\\
D=2N+1:&& \prod_{\alpha=0}^N A_\alpha = g^{-2N+1}\, c_N\, {\cal A}_{D-2}\, 
 \prod_i(8\pi J_i)\,,\nn
\eea
where $c_N=(-1)^{(N+1)/2}$, and ${\cal A}_{D-2}= 2 \pi^{(D-1)/2}/
\Gamma[(D-1)/2]$ is the volume of the unit $(D-2)$-sphere.

   The results presented above for black holes in gauged supergravities,
and for Kerr-AdS black holes in pure gravity with a cosmological constant,
admit straightforward limits to the ungauged, or zero cosmological constant,
case.  The radial functions in the metrics have a universal feature, as
can be seen in (\ref{5h}) for the example of five-dimensional gauged 
supergravity, that the degree of the polynomial in $r$ is reduced by 2
when the gauge coupling $g$ is set to zero.  In this
limit, the locations of these two "lost horizons" approach
$r=\pm \im\, g^{-1}$, and the areas of the lost horizons in the cases of
even and odd dimensional black holes are
\bea
D=2N+2:&& A_{\rm lost}= {(-1)^N\, g^{-2N}\,\cal A}_{D-2}\,,\\
D=2N+1:&& A_{\rm lost}=\mp\im\, (-1)^N\, g^{-2N+1}\, {\cal A}_{D-2}\,.\nn
\eea
If these areas are factored out from our previous expressions for the
horizon area products, and then $g$ is sent to zero, we can obtain
the analogous formulae for the corresponding ungauged supergravities,
and for asymptotically-flat rotatating black holes in arbitrary
dimensions.  For the black holes in four and five dimensional supergravities,
the limits yield expressions encompassed by those 
given above for the ungauged cases.  For the black holes in gauged
six and seven dimensional supergravities, it is interesting to note that
the electric charge terms scale to zero in the ungauged limit.
The resulting expressions are then just the $D=6$ and $D=7$ specialisations
of the limiting forms for asymptotically-flat black holes in 
arbitrary dimensions, which we find to be
\bea
D=2N+2:&& \quad \prod_{\alpha=1}^{2N}A_\alpha =
   \prod_{i=1}^N (8\pi J_i)^2\,,\nn\\
D=2N+1:&& \quad \prod_{\alpha=1}^N A_\alpha = 
 \prod_i(8\pi J_i)\,.\label{mpprod}
\eea
We have also worked out the area product formulae for a general class of
charged rotating black holes in $D>5$ ungauged supergravities \cite{CYD}, 
and we find the same phenomenon as in the $D=6$ and $D=7$ ungauged limits
described above.  Namely, the area products are independent of the charges
in $D>5$, and are given simply by the expressions (\ref{mpprod}) for 
uncharged asymptotically flat rotating black holes. 

   In this paper, we have obtained 
formulae for the products of the horizon areas in a wide variety of
black hole solutions, showing that they are independent of moduli and
are expressed solely in terms of quantised charges,
angular momenta and the gauge coupling constant.  These
provide tantalising hints of a possible explanation for the microscopic
properties of the black holes in terms of field theories in more than
two dimensions.
 
  We have not attempted here to address the question  
of whether these formulae remain  universal in the presence of 
external fields, as was done in certain four dimensional examples in 
\cite{Ansorg1,Ansorg2,Ansorg3,Ansorg4,Ansorg5,Ansorg6,Ansorg7}. 
This may be relatively straightforward in 
four and five dimensions, since the symmetries  allow a reduction to a system
of equations on a two-dimensional quotient space. We hope to return
to this subject in the future.
 The four-dimensional results in
\cite{Ansorg1,Ansorg2,Ansorg3,Ansorg4,Ansorg5,Ansorg6,Ansorg7}
are a promising indication that our quantisation results may be robust,
in the sense that they may survive in the presence of external fields. 
This is the analogue 
for black holes of the central  idea of Old Quantum Theory,
associated  with the names of Bohr, Wilson and Sommerfeld, 
that it is {\sl adiabatic invariants}  that should take quantised values 
because  classically their values do not change under slow 
perturbations.  

\noindent
{\bf Acknowledgements}

We are grateful to David Chow and Finn Larsen 
for useful discussions.  M.C. is supported in
part by DOE grant DE-FG05-95ER40893-A020, the 
Slovenian Agency for Research (ARRS) and the 
Fay R. and
Eugene L. Langberg Chair. C.N.P. is supported in part by DOE grant 
DE-FG03-95ER40917.

\nopagebreak

\end{document}